\documentstyle[11pt]{article}
\begin{document}
\begin{titlepage}
\begin{flushright}
BRX TH-344 \\
HEP-TH/9302113 \\
23 February 1993
\end{flushright}
\vspace{2cm}
\begin{center}
{\large Division Algebras, Galois Fields, Quadratic Residues} \\
\vspace{2cm}
{\large Geoffrey Dixon} \\
Department of Physics \\
Brandeis University \\
Waltham, MA 02254 \\
\vspace{1cm}
{\bf Abstract}
\end{center}
Intended for mathematical physicists interested in applications of the
division algebras to physics, this article highlights some of their more
elegant properties with connections to the theories of Galois fields and
quadratic residues.
\vspace{1cm}
\normalsize
\footnoterule
\noindent
{\footnotesize e-mail: Dixon@binah.cc.brandeis.edu}
\end{titlepage}
\newpage

The reals, {\bf R}, complexes, {\bf C}, quaternions, {\bf Q}, and octonions,
{\bf O}, are the normed
division algebras, proven by Hurwitz [1] to be the only ones of their kind.  My
interest in these
algebras arises from a faith I share with many mathematical physicists that
they are intimitely linked
to the design of our physical reality  [2,3] (and if they are not, well they
ought to be, and it
is a shame they are not).  In searching for the key to that link I have
encountered
many of the most beautiful properties of these algebras, including connections
to Galois theory and
to the theory of quadratic residue codes.  The former connections highlight the
elegant cyclic
multiplication rules of {\bf Q} and {\bf O}, and in combination with the latter
connections they
provide another explanation for the uniqueness of the collection.

The octonion algebra, {\bf O}, is often developed as an extension of the
quaternion algebra,
{\bf Q}.  Let $q_{i}$, i=1,2,3, be a conventional basis for the hypercomplex
quaternions.  These
elements associate, anticommute, and satisfy $q_{i}^{2}=-1$.  The
multiplication table for {\bf Q}
is then determined by
\begin{equation}
q_{i}q_{i+1} = q_{i+2} ,
\end{equation}
i=1,2,3, all indices modulo 3, from 1 to 3.

Relabel these quaternion units $e_{i}$, i=1,2,3, and introduce a new unit,
$e_{7}$, anticommuting with
each of the $e_{i}$, which satisfies $e_{7}^{2} = -1$.  Define three more
units:
\begin{equation}
e_{4}= e_{1}e_{7} \mbox{, } e_{5}=e_{2}e_{7} \mbox{, } e_{6}=e_{3}e_{7}.
\end{equation}
Let {\bf O} be the real algebra  generated from the $e_{a}$ a=1,...,7, such
that
$\{q_{1} \rightarrow e_{a}$, $q_{2} \rightarrow e_{b}$, $q_{3} \rightarrow
e_{c}\}$ defines an
injection of {\bf Q} into {\bf O} for (a,b,c)=(1,2,3), (1,7,4), (2,7,5),
(3,7,6), (1,6,5), (2,4,6),
(3,5,4).  Therefore, for example, $e_{1}(e_{7}e_{5})=e_{1}e_{2}=e_{3}
=-(-e_{3}) = -e_{4}e_{5} =
-(e_{1}e_{7})e_{5}$.  So unlike the complexes and quaternions, the octonions
are nonassociative.

Like {\bf C} and {\bf Q}, however, {\bf O} is a division algebra, and it is
normed.  In particular,
if $x=x^{0} + x^{a}e_{a}$, (sum a=1,...,7), and $x^{\dagger} = x^{0} -
x^{a}e_{a}$ (an
antiautomorphism), then
\begin{equation}
\Vert x \Vert ^{2} = x^{\dagger}x = \sum_{a=0}^{7} x^{a}x^{a}
\end{equation}
defines the square of the norm of $x$ (so $x^{-1}= x^{\dagger}/\Vert x
\Vert^{2}$).

This octonion multiplication is not, however, the most natural, and it will not
be employed in
here.  Again let $e_{a}, a=1,...,7$, represent the hypercomplex units, but now
adopt the cyclic
multiplication rule:
\begin{equation}
e_{a}e_{a+1}=e_{a+5},
\end{equation}
a=1,...,7, all indices modulo 7, from 1 to 7 (the right-hand side could be
changed to $e_{a+3}$,
which generates an alternative multiplication table for {\bf O}, dual to the
first in a sense outlined
below).  In particular,
\begin{equation}
\{q_{1} \rightarrow e_{a}, q_{2} \rightarrow e_{a+1}, q_{3} \rightarrow
e_{a+5}\}
\end{equation}
define injections of {\bf Q} into {\bf O} for a=1,...,7.  I am accustomed to
using the symbol $e_{0}$
to represent unity, and I bother to remember that although $7=0$ mod 7, $e_{7}
\ne e_{0}$, and in
the multiplication rule (4) the indices range from 1 to 7, and  the index $0$
is not subject to the
rule. (In [3] $\infty$ is used as the index for unity, and this has advantages,
which I find
intermittently persuasive.)

This octonion multiplication has some very nice properties.  For example,
\begin{equation}
\mbox{if  } e_{a}e_{b}=e_{c}, \mbox{  then  } e_{(2a)}e_{(2b)}=e_{(2c)}.
\end{equation}
(6) in combination with (4) immediately implies
\begin{eqnarray}
& e_{a}e_{a+2}=e_{a+3}, \nonumber \\
& e_{a}e_{a+4}=e_{a+6}
\end{eqnarray}
(so  $e_{a}e_{a+2^{n}}=e_{a-2^{n+1}}$, or
$e_{a}e_{a+b} = [b^{3} \mbox{ mod } 7]e_{a-2b^{4}}$, $b=1,...,6$, where $b^{3}$
out front provides the
sign of the product (modulo 7, $1^{3}=2^{3}=4^{3}=1$, and
$3^{3}=5^{3}=6^{3}=-1$ )).

These modulo 7 periodicity properties are reflected in the full multiplication
table:
\begin{equation}
\left[ \begin{array} {cccccccc}
1 & e_{1} & e_{2} & e_{3} & e_{4}  & e_{5} & e_{6} & e_{7}\\
e_{1}&-1&e_{6}&e_{4}&-e_{3}&e_{7}&-e_{2}&-e_{5}\\
e_{2}&-e_{6}&-1&e_{7}&e_{5}&-e_{4}&e_{1}&-e_{3}\\
e_{3}&-e_{4}&-e_{7}&-1&e_{1}&e_{6}&-e_{5}&e_{2}\\
e_{4}&e_{3}&-e_{5}&-e_{1}&-1&e_{2}&e_{7}&-e_{6}\\
e_{5}&-e_{7}&e_{4}&-e_{6}&-e_{2}&-1&e_{3}&e_{1}\\
e_{6}&e_{2}&-e_{1}&e_{5}&-e_{7}&-e_{3}&-1&e_{4}\\
e_{7}&e_{5}&e_{3}&-e_{2}&e_{6}&-e_{1}&-e_{4}&-1\\
\end{array} \right].
\end{equation}
The naturalness of this table is reflected in the matrix of its signs :
\begin{equation}
O = \left[ \begin{array} {cccccccc}
1&1&1&1&1&1&1&1\\
1&-1&1&1&-1&1&-1&-1\\
1&-1&-1&1&1&-1&1&-1\\
1&-1&-1&-1&1&1&-1&1\\
1&1&-1&-1&-1&1&1&-1\\
1&-1&1&-1&-1&-1&1&1\\
1&1&-1&1&-1&-1&-1&1\\
1&1&1&-1&1&-1&-1&-1\\
\end{array} \right].
\end{equation}
(This is what is called a normalized Hadamard matrix of order 8 [4].  It is
normalized because the
first row and column are all 1's, and it is a Hadamard matrix in containing
only 1's and -1's, and in
satisfying $OO^{\dagger} = 8I$, where $O^{\dagger}$ is the transpose of $O$,
and $I$ is the 8x8
identity matrix.)   Note that if $a \ne 0$, $b \ne 0$, then the components
$O_{a,b}=
O_{a+1,b+1}$, indices from 1 to 7, modulo 7 (first row and column of $O$ are
given the index
0).

Let $ O_{a}$ be the $a^{th}$ row of $O$, a=0,1,...,7, and define the product
\begin{equation}
 O_{a} \bullet O_{b} = O_{a,b}O_{c},
\end{equation}
where the components of $O_{c}$ are $O_{c,d}=O_{a,d} O_{b,d}$, for each
d=0,1,...,7, and $O_{a,b}$
gives a sign to the product.  For example, \\
\begin{eqnarray}
O _{1} \bullet O_{2} = +[1 \cdot 1,(-1) \cdot (-1), 1 \cdot (-1), 1 \cdot 1,
(-1) \cdot
1, 1 \cdot (-1), (-1) \cdot 1, (-1) \cdot (-1)]
\nonumber
\end{eqnarray}
\begin{eqnarray}
= +[1,1,-1,1,-1,-1,-1,1] = O_{6}, \nonumber
\end{eqnarray}
where the plus sign out front arises from the component ${\it O}_{1,2}=+1$.
The resulting
multiplication table of the ${\it O}_{a}$ is exactly the same as (8), giving
rise to the obvious
isomorphism $e_{a} \rightarrow {\it O}_{a}, a=0,1,...,7$.

The quaternion algebra arises in exactly the same way from the sign matrix
\begin{equation}
{\it Q} = \left[ \begin{array} {cccccccc}
1&1&1&1\\
1&-1&1&-1\\
1&-1&-1&1\\
1&1&-1&-1\\
\end{array} \right].
\end{equation}
Likewise the complexes arise from
\begin{equation}
{\it C} = \left[ \begin{array} {cccccccc}
1&1\\
1&-1\\
\end{array} \right].
\end{equation}
(These are normalized Hadamard matrices of order 4 and 2.)

The arrays used above are connected with Galois fields.  The real numbers are
the paradigm for
mathematical field theory.  There is addition (and subtraction), an additive
identity ,$0$, and
every element $x$ has an additive inverse ,$-x$.  There is multiplication (and
division), a
multiplicative identity ,$1$, and every element $x\ne 0$ has a multiplicative
inverse ,$x^{-1}$.
Multiplication by zero gives zero, and for all $x \ne 0$ and  $y \ne 0$, we
also have $xy \ne 0$ (no
divisors of zero).  Finally, $xy = yx$ (commutative), and $x(yz) = (xy)z$
(associative).

{\bf R} is an infinite field, but there also exist finite fields.  For any
prime $p$ there exist
(unique up to isomorphism) fields of order $p^{k}$ for all $k=1,2,3,...$,
denoted $GF(p^{k})$
($G$ for Galois, their ill-fated founder, $F$ for field).  For no other
positive integers are there
fields of that order.

The $p^{k}$ elements of $GF(p^{k})$ are easily written: $\{0, 1, h, h^{2}, ...,
h^{p^{k}-2}\}$.  That
is, the multiplication of $GF(p^{k})$ is cyclic and for all $x \ne 0$ in
$GF(p^{k})$,
\begin{equation}
x^{p^{k}-1} - 1 = 0
\end{equation}
(ie., $h^{p^{k}-1} = 1$).

All that remains then is to construct an addition table for $GF(p^{k})$
consistent with its being a
field.  This problem can be reduced to finding what is called a Galois sequence
for $GF(p^{k})$,
which consists of $p^{k}-1$ elements of $Z_{p}$ (the integers modulo $p$).  Its
further properties can
be best illustrated by an example.  (Mathematicians have a more elaborate
development in terms of
polynomials and quotient modules; the elements of a Galois sequence appear in
that context as
coefficients of a polynomial.)

$[\begin{array}{cccccccc}  0 & 1 & 1 & 2 & 0 & 2 & 2 & 1  \\ \end{array}]$ is a
Galois
sequence for $GF(3^{2}=9)$.  We identify it with $h^{0} = 1$, the
multiplicative identity of
$GF(9)$, and we'll identify its $k th$ cyclic permutation with $h^{k}$.  That
is,
\newpage
$$
h^{1} = [\begin{array}{cccccccc} 1 & 0 & 1 & 1 & 2 & 0 & 2 & 2 \\ \end{array}],
$$
$$
h^{2} = [\begin{array}{cccccccc} 2 & 1 & 0 & 1 & 1 & 2 & 0 & 2 \\ \end{array}],
$$
$$
h^{3} = [\begin{array}{cccccccc} 2 & 2 & 1 & 0 & 1 & 1 & 2 & 0 \\ \end{array}],
$$
$$
h^{4} = [\begin{array}{cccccccc} 0 & 2 & 2 & 1 & 0 & 1 & 1 & 2 \\ \end{array}],
$$
$$
h^{5} = [\begin{array}{cccccccc} 2 & 0 & 2 & 2 & 1 & 0 & 1 & 1 \\ \end{array}],
$$
$$
h^{6} = [\begin{array}{cccccccc} 1 & 2 & 0 & 2 & 2 & 1 & 0 & 1 \\ \end{array}],
$$
$$
h^{7} = [\begin{array}{cccccccc} 1 & 1 & 2 & 0 & 2 & 2 & 1 & 0 \\ \end{array}
       ],
$$
\begin{equation}
h^{8} = [\begin{array}{cccccccc} 0 & 1 & 1 & 2 & 0 & 2 & 2 & 1 \\ \end{array}],
\end{equation}
where $h^{8} = h^{0} = 1$ gets us back where we started (any cyclic permutation
of the
initial sequence would have been a valid starting point).  Notice that the
first $k=2$ elements of
each sequence are unique, and can be used as labels for the elements (we are
using instead the
exponents).  And notice that by adjoining to this collection the zero sequence,
$0 = [\begin{array}{cccccccc} 0 & 0 & 0 & 0 & 0 & 0 & 0 & 0 \\ \end{array}]$,
we have a set
of $p^{k} = 3^{2} = 9$ vectors (sequences), each $p^{k}-1 = 3^{2}-1 =
8$-dimensional over $Z_{p} =
Z_{3}$, and that the set is closed with respect to $Z_{3}$ vector addition.
For example, using
$+_{p}$ to represent modulo $p$ addition,
$$
h^{2} +_{3} h^{4} =
[\begin{array}{cccccccc} 2 & 1 & 0 & 1 & 1 & 2 & 0 & 2 \\ \end{array}] +_{3}
[\begin{array}{cccccccc} 0 & 2 & 2 & 1 & 0 & 1 & 1 & 2 \\ \end{array}]
$$
$$
= [\begin{array}{cccccccc} 2 & 0 & 2 & 2 & 1 & 0 & 1 & 1 \\ \end{array}] =
h^{5}.
$$
A full addition table for $GF(9)$ resulting from this sequence is listed below:
\begin{equation}
\begin{array}{ccccccccc}
0     & h^{1} & h^{2} & h^{3} & h^{4} & h^{5} & h^{6} & h^{7} & h^{8} \\
h^{1} & h^{5} & h^{8} & h^{4} & h^{6} & 0     & h^{3} & h^{2} & h^{7} \\
h^{2} & h^{8} & h^{6} & h^{1} & h^{5} & h^{7} & 0     & h^{4} & h^{3} \\
h^{3} & h^{4} & h^{1} & h^{7} & h^{2} & h^{6} & h^{8} & 0     & h^{5} \\
h^{4} & h^{6} & h^{5} & h^{2} & h^{8} & h^{3} & h^{7} & h^{1} & 0     \\
h^{5} & 0     & h^{7} & h^{6} & h^{3} & h^{1} & h^{4} & h^{8} & h^{2} \\
h^{6} & h^{3} & 0     & h^{8} & h^{7} & h^{4} & h^{2} & h^{5} & h^{1} \\
h^{7} & h^{2} & h^{4} & 0     & h^{1} & h^{8} & h^{3} & h^{3} & h^{6} \\
h^{8} & h^{7} & h^{3} & h^{5} & 0     & h^{2} & h^{1} & h^{4} & h^{4} \\
\end{array}
\end{equation}
(recall that $h^{8}=1$).    Note that $h^{k}+_{3}h^{k} = h^{k+4}$ and
$h^{k}+_{3}h^{k}+_{3}h^{k} =
h^{k}+_{3} h^{k+4} = 0$.  Also, $h^{k}+_{3}h^{k+1} = h^{k+7}$.  Because for any
$x$ and
$y$ in any $GF(3^{m})$,
\begin{equation}
(x+_{3}y)^{3} = x^{3} +_{3} y^{3},
\end{equation}
cubing the last equation above results in $h^{k}+_{3}h^{k+3} = h^{k+5}$
(exponents are taken modulo
8 from 1 to 8, and although strictly speaking the exponents $k$ cube to $3k$,
because 3 and 8 are
relatively prime we are allowed to replace $3k$ by $k$ in constructing new
addition rules), and cubing
this leads back to $h^{k}+_{3}h^{k+1} = h^{k+7}$.  There is also,
$h^{k}+_{3}h^{k+2} = h^{k+3}$, which cubed yields, $h^{k}+_{3}h^{k+6} =
h^{k+1}$, and also
$h^{k}+_{3}h^{k+5} = h^{k+2}$, which cubed yields, $h^{k}+_{3}h^{k+7} =
h^{k+6}$.

Of more interest to us here are the fields $GF(2^{n}), n=1,2,3$.  In
particular, a Galois sequence
for $GF(2^{1})$ is $[\begin{array}{c}   1  \\ \end{array}]$,
for $GF(2^{2})$ is $[\begin{array}{ccc}  0 &  1 & 1 \\ \end{array}]$, and
for $GF(2^{3})$ is $[\begin{array}{ccccccc} 0 & 0 & 1 & 0 & 1 & 1 & 1  \\
\end{array}]$.
In this last case we define
$$
e^{1} = [\begin{array}{ccccccc}  1 & 0 & 0 & 1 & 0 & 1 & 1 \\ \end{array}],
$$
$$
e^{2} = [\begin{array}{ccccccc}  1 & 1 & 0 & 0 & 1 & 0 & 1 \\ \end{array}],
$$
$$
e^{3} = [\begin{array}{ccccccc}  1 & 1 & 1 & 0 & 0 & 1 & 0 \\ \end{array}],
$$
$$
e^{4} = [\begin{array}{ccccccc}  0 & 1 & 1 & 1 & 0 & 0 & 1 \\ \end{array}],
$$
$$
e^{5} = [\begin{array}{ccccccc}  1 & 0 & 1 & 1 & 1 & 0 & 0 \\ \end{array}],
$$
$$
e^{6} = [\begin{array}{ccccccc}  0 & 1 & 0 & 1 & 1 & 1 & 0 \\ \end{array}],
$$
\begin{equation}
e^{7} = [\begin{array}{ccccccc}  0 & 0 & 1 & 0 & 1 & 1 & 1 \\ \end{array}].
\end{equation}
Addition in this case can also be completely described by cyclic equations in
the $e^{a}$.  To begin
with,
\begin{equation}
e^{a} +_{2} e^{a} = 0
\end{equation}
(every element is its own additive inverse).  Also,
\begin{equation}
e^{a} +_{2} e^{a+1} = e^{a+5}.
\end{equation}
Since in the $p=2$  case
\begin{equation}
(x +_{2} y )^{2} = x^{2} +_{2} y^{2},
\end{equation}
squaring the above addition rule leads to a new rule,
\begin{equation}
e^{a} +_{2} e^{a+2} = e^{a+3},
\end{equation}
and squaring this leads to
\begin{equation}
e^{a} +_{2} e^{a+4} = e^{a+6}
\end{equation}
(exponents are taken modulo 7 from 1 to 7).

The link of $GF(8)$ to the octonions should now be obvious.  The matrix of
signs in (9), used to
construct an octonion multiplication, could have been replaced by the following
matrix of elements
of $Z_{2}$ (ie., $0$'s and $1$'s):
\begin{equation}
O' = \left[\begin{array}{cccccccc}
 0 & 0 & 0 & 0 & 0 & 0 & 0 & 0  \\
 0 & 1 & 0 & 0 & 1 & 0 & 1 & 1  \\
 0 & 1 & 1 & 0 & 0 & 1 & 0 & 1  \\
 0 & 1 & 1 & 1 & 0 & 0 & 1 & 0  \\
 0 & 0 & 1 & 1 & 1 & 0 & 0 & 1  \\
 0 & 1 & 0 & 1 & 1 & 1 & 0 & 0  \\
 0 & 0 & 1 & 0 & 1 & 1 & 1 & 0  \\
 0 & 0 & 0 & 1 & 0 & 1 & 1 & 1  \\
\end{array}\right].
\end{equation}
Note that $(-1)^{O'_{ab}} = O_{ab}$ (see (9)), so if we define
\begin{equation}
O'_{a}*O'_{b} = (-1)^{O'_{ab}}[O'_{a} +_{2} O'_{b}],
\end{equation}
then we have once again created an octonion product, where this time the rows
of $O'$ are identified
with the basis of the octonions.  Note!  We have used $GF(8)$ addition to
create an octonion
multiplication.  The first row of $O'$ is the multiplicative identity of {\bf
O}, and we must create
a new 0 to play the role of the additive identity of {\bf O}.  With respect to
{\bf O} addition, the
rows of $O'$ are now treated as linearly independent, a basis for a real
algebra.

Relabel the rows of $O'$ as $e_{a}, a=0,1,...,7$.  So the exponents of $GF(8)$
in (17) are mapped
into the subscripts of the octonions.  Because the octonion product (now
denoted just $e_{a}e_{b}$) is
derived directly from the $GF(8)$ addition, the exponent rules (19,21,22) are
valid for the octonion
product, the rules now applied to subscripts (see (4,7)).  In addition, the
index doubling
automorphism for the octonions (6) is now seen to follow from (20).

[Note: The sum rules (19,21,22) for $GF(8)$ correspond to (4,7), but in general
we can only make
such correspondences up to a sign.  For example, while it is true in $GF(8)$
that
$e^{a}+_{2}e^{a+5}=e^{a+1}$, in {\bf O} we have $e_{a}e_{a+5}=-e_{a+1}$.  Index
doubling is also
tricky, and in {\bf Q} it works out slightly differently.]

[Also note: In $GF(8)$, $e^{7}=e^{0}=1$.  The reason that it was listed as
$e^{7}$ in (17) is to
make the correspondence $e^{7} \rightarrow e_{7}$ of $GF(8)$ to {\bf O}.
Therefore, since
$e^{0}=e^{7}$, we have $e^{0} \rightarrow e_{7}$, too!  That is,  $e_{0} = 1$
has no
correspondence to any power of $e^{1} \in GF(8)$.  At this point the the
notation $e_{\infty} = 1$
becomes increasingly attractive. ]

[Finally note: the transpose of $O'$ also results in a valid $GF(8)$ addition
and {\bf O}
multiplication.  In this case, however, $e_{a}e_{a+1}=-e_{a+3}$ in {\bf O}.
Except for the sign
change, this is the dual multiplication mentioned above.  If we replace (24) by
$$
O'_{a}*O'_{b} = (-1)^{O'_{ba}}[O'_{a} +_{2} O'_{b}],
$$
we generate the {\bf O} multiplication rule,
$e_{a}e_{a+1}=-e_{a+5}$; and if we use the transpose of $O'$, the rule
$e_{a}e_{a+1}=e_{a+3}$.]

Having made the correspondence between $GF(8)$ addition and {\bf O}
multiplication, one is naturally
led to consider the role of $GF(8)$ multiplication in {\bf O}.  Since in
$GF(8)$,
$e^{a}e^{b}=e^{a+b}$, this operation on the indices of {\bf O} is just a cyclic
shift (of the index
$a$ for $a=1,...,7$; $e_{0}$ is left unaltered).   Let $S$ be the {\bf O}
automorphism that shifts the
indices of $e_{b}, b=1,...,7$ by $1$.  So $S^{a}$ shifts the {\bf O} indices by
$a$, and
$S^{7}=S^{0}$ is the identity map.  Let $\phi$ be the zero map, mapping all
$x\in {\bf O}$ to $0$.
Obviously this collection of eight maps can be made into the field $GF(8)$ if
given the appropriate
addition.  This may or may not be of interest, but this is as far down that
road as I am willing to
go at present.

In the quaternion case one makes a correspondence with $GF(4)$.  Everything
works out much the same,
save that (20) doesn't give rise to as simple a relation in {\bf Q} as it did
in {\bf O}.  By
inspection we see in this case that
\begin{equation}
\mbox{if } q_{i}q_{j}=q_{k}, \mbox{ then } q_{(2i)}q_{(2j)}=-q_{(2k)}.
\end{equation}
Index quadrupalling gets us back to $q_{i}q_{j}=q_{k}$, since 4=1 mod 3.  Hence
in {\bf O},
$e_{a}e_{b}=e_{c}$ could not imply $e_{(2a)}e_{(2b)}=-e_{(2c)}$, since
$2^{3}=8=1$ mod 7,  and
three (an odd number of) applications of index doubling must get us back to
$e_{a}e_{b}=e_{c}$.

The binary matrix generating both {\bf Q} multiplication and $GF(4)$ addition
is
\begin{equation}
Q' = \left[\begin{array}{cccc}
 0 & 0 & 0 & 0   \\
 0 & 1 & 0 & 1   \\
 0 & 1 & 1 & 0   \\
 0 & 0 & 1 & 1  \\
\end{array}\right].
\end{equation}
In both $O'$  and $Q'$, the first row of each after the zeroth must be either
the one
shown, or the first row of the respective transposes, for algebras isomorphic
to {\bf O} and {\bf Q}
to result from the process outlined.  In particular, consider
\begin{equation}
B = \left[\begin{array}{cccc}
 0 & 0 & 0 & 0   \\
 0 & 0 & 1 & 1  \\
 0 & 1 & 0 & 1   \\
 0 & 1 & 1 & 0   \\
\end{array}\right].
\end{equation}
[$B_{11}$ $B_{12}$ $B_{13}$] = [0 1 1] is also a Galois sequence for $GF(4)$,
but in this case the
algebra multiplication
\begin{equation}
B_{i}*B_{j} = (-1)^{B_{ij}}[B_{i} +_{2} B_{j}]
\end{equation}
does not result in {\bf Q}, but rather an algebra isomorphic to that generated
by the adjoint
elements, $q_{L1}q_{R3}$, $q_{L2}q_{R2}$, $q_{L3}q_{R1}$.  Here the subscripts
$L$ and $R$ denote
multiplication from the left and right on {\bf Q}.  Since $q_{Li}q_{Rj}[x] =
q_{i}xq_{j} =
q_{Rj}q_{Li}[x]$, it is apparent left  adjoint multiplication commutes with
right.  (This is not the
case for {\bf O}, which is complicated by nonassociativity  [2,5].)

Addition on $GF(2^{n})$ can be turned into an algebra multiplication in the way
outlined
for $n > 3$ as well.  For example, let
$$
g^{1} = [\begin{array}{ccccccccccccccc}
1 & 0 & 0 & 0 & 1 & 0 & 0 & 1 & 1 & 0 & 1 & 0 & 1 & 1 & 1\\ \end{array}],
$$
15-dimensional over $Z_{2}$.  This is a Galois sequence for $GF(16)$, and it
can be used to construct
a new 16-dimensional algebra, extending the sequence, {\bf R}, {\bf C}, {\bf
Q}, {\bf O} (this is
distinct from the Cayley-Dickson prescription, which is founded on the
inclusion property, and in
fact  {\bf O} is not a subalgebra of this new 16-dimensional algebra, which is
noncommutative,
nonassociative, and nonalternative; in [6] binary sequences are used to
construct the
Cayley-Dickson multiplication rules, as well as those of Clifford algebras).

One final path down which I have no intention of travelling far: we should be
able to construct
algebras in like manner from any $GF(p^{n})$, for any prime $p$.  For example,
take the $h^{k},
k=1,...,8$, in $GF(9)$ listed in (14), and map them to $h_{k}, k=1,...,8$, part
of a basis for a
new algebra.  Map the zero sequence to $1$, completing the basis.  Form the
stacked sequences in
(14) into a matrix, $H$ ($8 \times 8$).  If $h^{i}+_{3}h^{j}=h^{k}$ in $GF(9)$,
then define
\begin{equation}
h_{i}h_{j}=exp[2\pi iH_{ij}/3]h_{k}.
\end{equation}
If $j-i=4$ mod 8, then replace $h_{k}$ by 1.  Here we have yet another algebra,
but at this point I'm
just spewing out ideas without a clear notion of their interest or viability,
so I'll shift directions
a bit in hopes of bringing order out of chaos.

It would seem in light of the material presented to this point that the
division
algebras are four out of an infinite collection of possible algebras
constructable in like manner.
And it {\it is} a collection, not a sequence.  Highlighting this is the fact
that the first rows
of $Q'$ and $O'$ (ignoring the intitial $0$'s) had to be
$[\begin{array}{ccc} 1 & 0 & 1 \\ \end{array}]$ and
$[\begin{array}{ccccccc} 1 & 0 & 0 & 1 & 0 & 1 & 1  \\ \end{array}]$ for {\bf
Q} and {\bf O} with
the multiplication rules we are adopting to result (see (27,28)).  Completely
different algebras
result from most of the other cyclic permutations of these sequences.

We could also have begun with the dual sequences (beginning with the same
element, but in reverse
order), $[\begin{array}{ccc} 1 & 1 & 0 \\ \end{array}]$ and
$[\begin{array}{ccccccc} 1 & 1 & 1 & 0 & 1 & 0 & 0 \\ \end{array}]$.  These
sequences also give
rise to {\bf Q} and {\bf O}, and they are Galois sequences for $GF(4)$ and
$GF(8)$.  They are in
addition {\it quadratic residue codes} of lengths 3 and 7 over $GF(2)$ [4].
For
example, the quadratic residues modulo 7 are  $0^{2} = 7^{2} = 0$, $1^{2} =
6^{2} = 1$, $2^{2} =
5^{2} = 4$, $3^{2} = 4^{2} = 2$, so confusingly renumbering the positions of
the sequence above 0 to
6, we see that the 1's appear in the 0, 1, 2, and 4 positions, which are
determined by the quadratic
residues.  Likewise, modulo 3, $0^{2} = 3^{2} = 0$, $1^{2} = 2^{2} = 1$, and
the 1's of
$[\begin{array}{ccc} 1 & 1 & 0 \\ \end{array}]$ appear in the 0 and 1
positions.  The quadratic
residue code of length 1 over $GF(2)$  is $[\begin{array}{c} 1  \\
\end{array}]$, also the Galois
sequence of $GF{2}$, and associated with {\bf C}.

There are {\it no} other examples of quadratic residue codes of any prime
length $p$ over $GF(2)$ that
correspond to Galois sequences.  To even have a chance we must have a code of
length $2^{k}-1$, and
$2^{k}-1$ must be prime.  So 15 is out.  The quadratic residue code of length
31 is
$$
[1110110111100010101110000100100],
$$
and a Galois sequence, equal to (29) in the first 7 places, is
$$
[1110110001111100110100100001010].
$$
Let $U$ be the 31x31 matrix formed of the first of these sequences and all its
cyclic permutations,
and let $V$ be the 31x31 matrix formed from the second.  The first has the nice
property shared by
all quadratic residue codes over $GF(2)$ that
\begin{equation}
(-1)^{U_{ab}} = -(-1)^{U_{ba}}, a \ne b.
\end{equation}
In the $2^{2}-1=3$ and  $2^{3}-1=7$ cases this gives rise to the
noncommutativity among the imaginary
basis elements ($\ne 1$) of {\bf Q} and  {\bf O}, which together with
\begin{equation}
(-1)^{U_{aa}} = -1
\end{equation}
ensures that {\bf Q} and {\bf O} are division algebras (replace $U$ by the
appropriate 3x3 and 7x7
matrices).  But unfortunately the rows of $U$ are not closed under $Z_{2}$
addition.  Those of $V$
are are closed under $Z_{2}$ addition.  For all $a,b \in \{1,...,31\}$, $a \ne
b$,
\begin{equation}
V_{a} +_{2} V_{b} = V_{c},
\end{equation}
for some $c$.  So $V$ gives rise to an algebra, but because
\begin{equation}
(-1^{V_{ab}} \ne -(-1)^{V_{ba}}, a\ne b,
\end{equation}
in general, there will be divisors of zero, and the algebra is not a division
algebra.
Requiring of our generating sequences that they be both Galois and quadratic
residue is a heavy
restriction, and the division algebras are the only algebras that result.

The quaternion and octonion codes/Galois sequences arise in other contexts.
For example, they are
useful in constructing the special lattices $D_{4}$ and $E_{8}$ [4] associated
with the integral
quaternions and octonions, and they arise in connection with projective
geometry [7].

Finally, it is my belief that the laws of Nature will be found to accrete about
the most special,
select, and generative of mathematical objects and ideas (a kind of
hypervariational principle) that
spawned my interest in the division algebras.  At the very least it can not be
doubted they are
special, select, and generative. \\ \\
{\bf Acknowledgement} \\
I would like to thank Paul Monsky for information on Galois sequences. \\ \\
{\bf References:} \\ \\
1. A. Hurwitz, {\it \"Uber die Composition der quadratischen Formen von
beliebig vielen Variablen}, Nachrichten von der Gesellschaft der
Wissenschaften zu G\"ottingen, 309 (1898). \\
2. G.M. Dixon, {\it Derivation of the Standard Model}, Il Nuovo
Cimento {\bf 105B}, 349(1990). \\
3. M. G\"unaydin and F. G\"ursey, Phys. Rev. D {\bf 10}, 674(1983). \\
P. Goddard, W. Nahm, D.I. Olive, H. Ruegg and A. Schwimmer,
Comm. Math. Phys. {\bf 112}, 385(1987). \\
F. Smith, {\it Hermitian Jordan Triple Systems, the Standard Model
plus Gravity, and $\alpha_{E} = 1/137.03608$}, hep-th 9302030. \\
4. J.H. Conway, N.J.A. Sloane, "Sphere Packings, Lattices and Groups",
Springer-Verlag, 2nd edition, 1991. \\
5. C.A. Manogue, J. Schray, {\it Finite Lorentz Transformations,
Automorphisms, and Division Algebras}, hep-th 9302044. \\
6. P-E. Hagmark and P. Lounesto, {\it Walsh Functions, Clifford
Algebras, and Cayley-Dickson Process}, in "Clifford Algebras and
Their Applications in Mathematical Physics", D. Reidel Publishing
Company, 531(1986). \\
7. R. Shaw, {\it Symmetry}, "Mathematical Perspectives: Four
Recent Inaugural Lectures", Hull University Press, 77(1991).

\end{document}